\documentclass{iau}
\usepackage{graphicx}

\title[Star Formation in z$\sim$6 Quasars] 
{Star Formation in Quasar Host Galaxies at Redshift 6: 
Millimeter Surveys and New Insights from ALMA}

\author[Wang et al.]   
{Ran Wang$^{1,2,11}$, Jeff Wagg$^{3}$, Chris L. Carilli$^1$, 
 Fabian Walter$^4$, Xiaohui Fan$^2$, 
 Frank Bertoldi$^5$
 Dominik. A. Riechers$^{6}$,
 Alain Omont$^7$,
 Karl M. Menten$^8$,
 Pierre Cox$^9$,
 Michael A. Strauss$^{10}$,
 Desika Narayanan$^2$}

\affiliation{$^1$National Radio Astronomy Observatory, PO Box 0, Socorro, NM, 
USA 87801, \\ email: {\tt rwang@nrao.edu} \\[\affilskip] 
$^2$Steward Observatory, University of Arizona, 933 N Cherry Ave., Tucson, AZ, 85721, USA 
$^3$European Southern Observatory, Alonso de C\'ordova 3107, Vitacura, 
Casilla 19001, Santiago 19, Chile 
$^4$Max-Planck-Institute for Astronomy, K$\rm \ddot o$nigsstuhl 17, 
69117 Heidelberg, Germany 
$^5$Argelander-Institut f$\rm \ddot u$r Astronomie,
University of Bonn, Auf dem H$\rm \ddot u$gel 71, 53121 Bonn, Germany
$^6$Astronomy Department, Cornell University, 220 Space Sciences Building,
Ithaca, NY 14853, USA 
$^7$Institut d'Astrophysique
de Paris, CNRS and Universite Pierre et Marie Curie, Paris, France
$^8$Max-Planck-Institut f$\rm \ddot u$r Radioastronomie, Auf dem 
H$\rm \ddot u$gel 71, 53121 Bonn, Germany 
$^9$Institute de Radioastronomie Millimetrique, St. Martin d'Heres, F-38406, France 
$^{10}$Department of Astrophysical Sciences, Princeton University, Princeton, NJ, 
USA, 08544 \\[\affilskip] $^{11}$Jansky Fellow}

\pubyear{2012}
\volume{292}  
\pagerange{xxx--xxx}
\jname{Molecular Gas, Dust, and Star Formation in Galaxies}
\editors{Tony Wong \& Juergen Ott}
\begin{document}

\maketitle

\begin{abstract}
We have been carrying out a systematic survey of the star formation and ISM properties in
the host galaxies of z$\sim$6 quasars. Our 250 GHz observations,  
together with available data from the literature, yield 
a sample of 14 z$\sim$6 quasars that are bright in millimeter dust 
continuum emission with estimated FIR luminosities of a few $\rm 10^{12}$ 
to $\rm 10^{13}\,L_{\odot}$.
Most of these millimeter-detected z$\sim$6 quasars have also been detected 
in molecular CO line emission, indicating molecular gas masses on order 
of $\rm 10^{10}\,M_{\odot}$. We have searched for [C II] 158 micron fine structure 
line emission toward four of the millimeter bright z$\sim$6 quasars with ALMA 
and all of them have been detected. All these results suggest massive  
star formation at rates of about 600 to 2000 $\rm M_{\odot}\,yr^{-1}$
over the central few kpc region of these quasar host galaxies.  
\end{abstract}

\firstsection 
\section{Introduction}
More than fifty quasars at $\sim$6 have 
been detected from large optical and near-infrared surveys, such as 
the Sloan Digital Sky Survey (hereafter SDSS, e.g., \cite[Fan et al. 2006]{fan06}) 
and the Canada-France High redshift Quasar Survey (CFHQS, \cite[Willott et al. 
2007]{willott07}). These objects represent the first generation
of supermassive black holes (SMBHs) which formed within 1 Gyr of the Big Bang and are
accreting at their Eddington limit (e.g. \cite[Kurk et al. 2007]{kurk07}).
Strong millimeter dust continuum, molecular CO, and [C II] 158 $\mu$m fine structure 
line emission were first detected from the z=6.42 quasar 
SDSS J114816.64+525150.3 (hereafter J1148+5251, \cite[Bertoldi et al. 2003a]{bertoldi03a}; 
\cite[Maiolino et al. 2005, 2012]{maiolino05,maiolino12}; \cite[Walter et al. 2003]{walter03}; 
\cite[Riechers et al. 2009]{riechers09}), indicating intense 
star formation with a peak surface density of $\rm \sim1000\,M_{\odot}\,yr^{-1}\,kpc^{-2}$ 
over the central 1.5 kpc region (\cite[Walter et al. 2009]{walter09}). 
The dynamical mass estimated with the resolved CO line emission indicates 
SMBH-bulge mass ratios more than one order
of magnitude higher than the typical value found in the present 
universe (\cite[Walter et al. 2004]{walter04}).
The millimeter observations of J1148+5251 suggest an early phase 
of SMBH-galaxy evolution and the dust continuum, molecular/atomic 
line emission have been searched in more quasars at the 
highest redshift (e.g. \cite[Carilli et al. 2007]{carilli07}; 
\cite[Wang et al. 2008, 2011a, 2011b]{wang08, wang11a, wang11b}; 
\cite[Venemans et al. 2012]{Venemans12}).
In this paper, we summarize recent millimeter observations 
of the sample of quasars known at z$\sim$6 and discuss the 
star forming activity in these earliest quasar host galaxies.

\section{Millimeter observations of the z$\sim$6 quasars}

A sample of 41 quasars at z$\sim$6 have been observed at 250 GHz 
using the MAMBO bolometer array on the IRAM 30-m telescope 
(\cite[Bertoldi et al. 2003a]{bertoldi03a}; \cite[Petric et al. 2003]{petric03};
\cite[Willott et al. 2007]{willott07}; \cite[Wang et al. 2008, 2011a]{wang08,wang11a}). 
This 250 GHz observed sample includes all the most luminous (i.e. SDSS z-band 
magnitudes of $\rm 18.74\leq z_{AB}\leq20.42$) 
objects from the SDSS main survey 
(e.g. \cite[Fan et al. 2006]{fan06}) and the 
objects from the SDSS southern deep imaging survey and 
the CFHQS (e.g. \cite[Jiang et al. 2009]{jiang09}; 
\cite[Willott et al. 2007]{willott07}) that are one to two magnitudes fainter 
in the optical compared to the SDSS main survey. The MAMBO observations 
have reached a typical 1$\sigma$ sensitivity of $\rm \sim0.6\,mJy$ and 
14 out of the 41 objects are detected at $\rm \geq3\sigma$, yielding a 
detection rate of $\rm 34\pm9$\%. This is consistent with the (sub)mm 
detection rates of optically selected quasars at redshifts 2 
and 4 (\cite[Priddey et al. 2003]{priddey03}; \cite[Omont et al. 2001, 
2003]{omont01,omont03}; \cite[Carilli et al. 2001]{carilli01}). 

The FIR luminosities estimated with the 250 GHz flux densities for 
the millimeter-detected z$\sim$6 quasars are a few $\rm 10^{12}$ 
to $\rm 10^{13}\,L_{\odot}$(\cite[Wang et al. 2008, 2011]{wang08,wang11}). 
In the left panel of Figure 1, we compare the FIR-to-AGN 
bolometric luminosity relation of the z$\sim$6 quasars to the local 
optically selected PG quasars and a sample of IR luminous type I quasars 
hosted in ultra-luminous infrared galaxies (\cite[Hao et al. 2005]{hao05}). 
Most of the  millimeter-detected 
z$\sim$6 quasars follow the shallower luminosity correlation trend 
defined by the IR quasars (\cite[Wang et al. 2008, 2011]{wang08,wang11}). 
This may suggest a starburst-dominant 
FIR emission in the millimeter bright quasars at z$\sim$6,
similar to that found in the local IR quasars. 
We also calculate the average FIR emission by stacking the 250 GHz measurements 
for the MAMBO non-detections in two quasar luminosity bins (i.e. optically bright, 
rest-frame 1450 $\rm \AA$ AB magnitude $\rm m_{1450}<20.2$, 
and optically faint, $\rm m_{1450}\geq20.2$, see \cite[Wang et al. 2011]{wang11}). 
The average FIR luminosity/upper limit for the non-detections is consistent with 
the trend defined by the local PG quasars. 

Molecular CO (6-5) line emission has been detected in eleven of the 
millimeter-detected quasars at z$\sim$6 using the IRAM Plateau de Bure 
Interferometer (PdBI). The result 
indicates highly excited molecular gas on order of $\rm 10^{10}\,M_{\odot}$ 
in the quasar host galaxies (\cite[Bertoldi et al. 2003b]{bertoldi03b}, 
\cite[Riechers et al. 2009]{riechers09}; \cite[Carilli et al. 2007]{carilli07}; 
\cite[Wang et al. 2010, 2011]{wang10, wang11}). The FIR and CO luminosities 
follow the relationship defined by actively star-forming
galaxies at low and high redshifts (\cite[Riechers et al. 2006]{riechers06}; 
\cite[Wang et al. 2010]{wang10}).

We are also searching for [C II] 158 micron fine 
structure line emission in our quasar sample. We have an on-going ALMA 
Cycle 0 program to look for this line in z$\sim$6 quasars 
with 250 GHz continuum detections (Wang et al. 2012, in prep.). The data have been obtained
for four of them in the extended configuration with typical resolution 
of $\rm \sim 0.7''$ and all the four objects show clear detections. 
The [C II] luminosities are about 1 to
8$\rm \times10^{9}\,L_{\odot}$ and the [C II]-FIR luminosity ratios
are of the order of $\rm 10^{-4}$, values that are comparable to those found for other high-z
[C II]-detected quasars and about one order of magnitude lower than the
typical value of star forming galaxies (e.g. \cite[Maiolino et al. 2009]{maiolino09}; 
\cite[Stacey et al. 2010]{stacey10}; \cite[Wagg et al. 2012]{wagg12}). 
The line velocity maps of three of them show indications of 
velocity gradients along the major axis direction. We plot the line intensity 
and velocity map for one of the detections, ULAS J131911.29+095051.4 at z=6.132 
(\cite[Mortlock et al. 2009]{mortlock09}), in the right panel of Figure 1. The line emission is marginally resolved by
the $\rm 0.7''\times0.5''$ beam, with a deconvolved size of
$\rm 0.6''\times0.3''$ (i.e. $\rm 3.5\,kpc\times1.7\,kpc$, fitted to a two-dimensional
Gaussian distribution). 

\section{Discussion: star formation in the millimeter bright z$\sim$6 quasars}

The detections of stong dust continuum, molecular CO and [C II] fine 
structure line emission from the millimeter bright z$\sim$6 quasars 
strongly suggest active star formation in their host galaxies. 
In particular, the [C II] detections and the line velocity maps from 
our ALMA observation suggest a nuclear starburst disk over the central 
few kpc region. If we conservatively assume 
that 50\% of the FIR emission is powered by host galaxy star
formation, the estimated star formation rates (SFR) in the quasar host 
galaxies are about 600 
to 2000 $\rm M_{\odot}\,yr^{-1}$ (adopting a standard Salpeter initial mass
function, \cite[Kennicutt 1998]{kennicutt98}). These together with the molecular gas masses 
measured from the CO line emission yield gas depletion time scales
of $\rm \tau_{dep}=M_{gas}/SFR\sim1$--$\rm 3\times10^{7}\,yr$. 

\section{Summary}

We summarize recent millimeter observations of the sample of 
quasars at z$\sim$6. About 30\% of these objects have been 
detected in strong 250 GHz dust continuum and molecular CO line 
emission, and our ongoing ALMA observations also detected 
bright [C II] fine structure line emission in the central 
few kpc region of the quasar host galaxies. The results suggest 
massive star formation in the quasar host galaxies, which is 
in good agreement with the picture of supermassive black hole-galaxy co-evolution
at their earliest evolutionary epoch. Further high-resolution imaging 
of the dust, molecular, and atomic line emission in these objects 
(e.g. with Cycle 1 and the full configuration of ALMA) will fully 
probe the gas distribution, star formation rate surface density, star formation 
efficiency, and dynamical properties of the spheroidal quasar 
stellar bulges and address the SMBH-bulge relationships and quasar-galaxy 
evolution at the highest redshift. 

This work is based on observations carried out with the Max Planck Millimeter
Bolometer Array (MAMBO) on the IRAM 30m telescope, the Plateau
de Bure Interferometer, and ALMA (NRAO). IRAM is supported
by INSU/CNRS (France), MPG (Germany) and IGN (Spain). The National
Radio Astronomy Observatory (NRAO) is a facility of the National
Science Foundation operated under cooperative agreement by Associated
Universities, Inc. This paper makes use of the following ALMA data: 
ADS/JAO.ALMA\# 2011.0.00206.S . ALMA is a partnership 
of ESO (representing its member states), NSF (USA) and NINS (Japan), 
together with NRC (Canada) and NSC and ASIAA (Taiwan), in 
cooperation with the Republic of Chile. The Joint ALMA Observatory 
is operated by ESO, AUI/NRAO and NAOJ.

\begin{figure}[h]
\hspace*{-0.5in}
\includegraphics[width=3.7in]{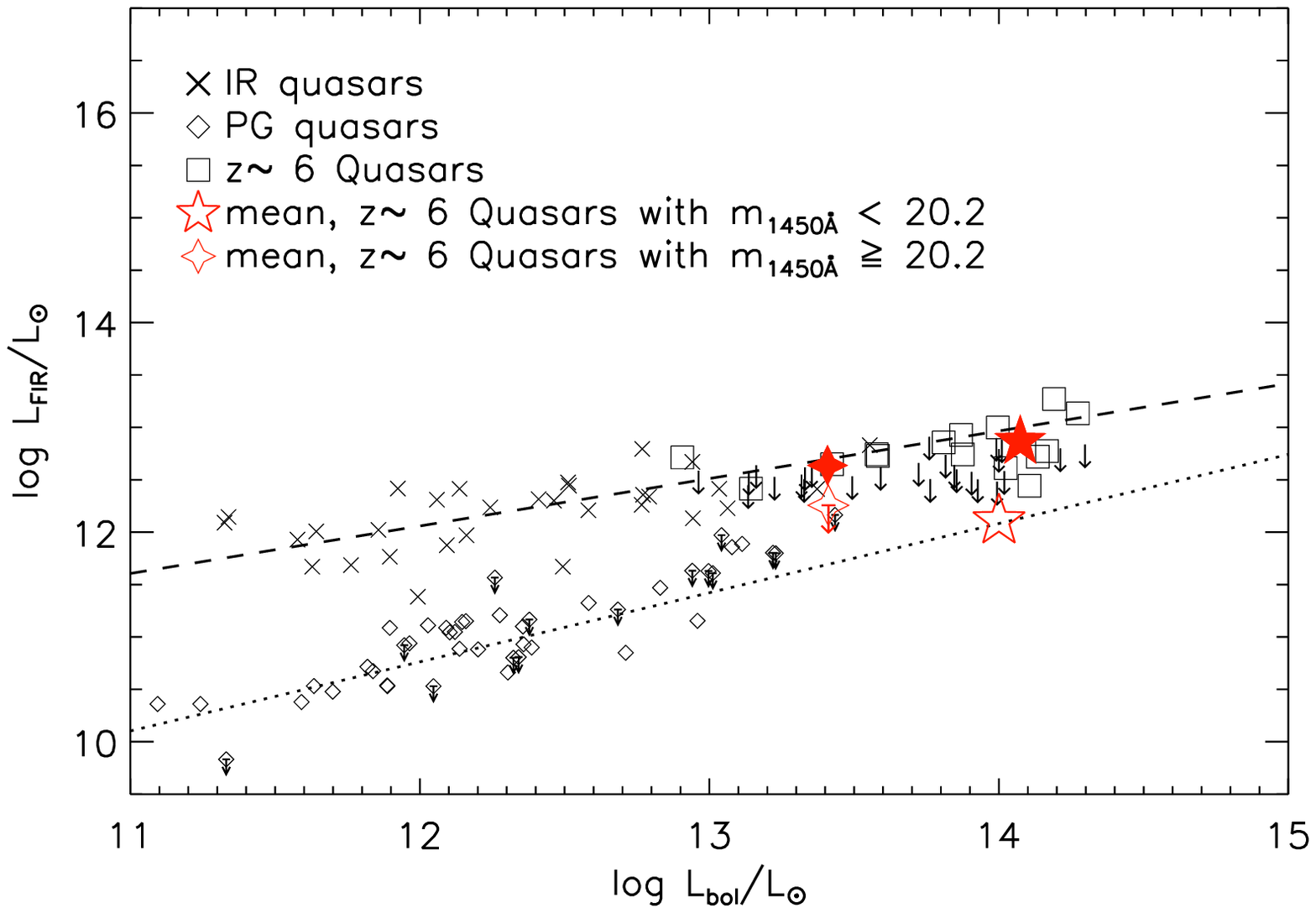}
\vskip -2.5in
\hspace*{3.5in}
\includegraphics[width=3.0in]{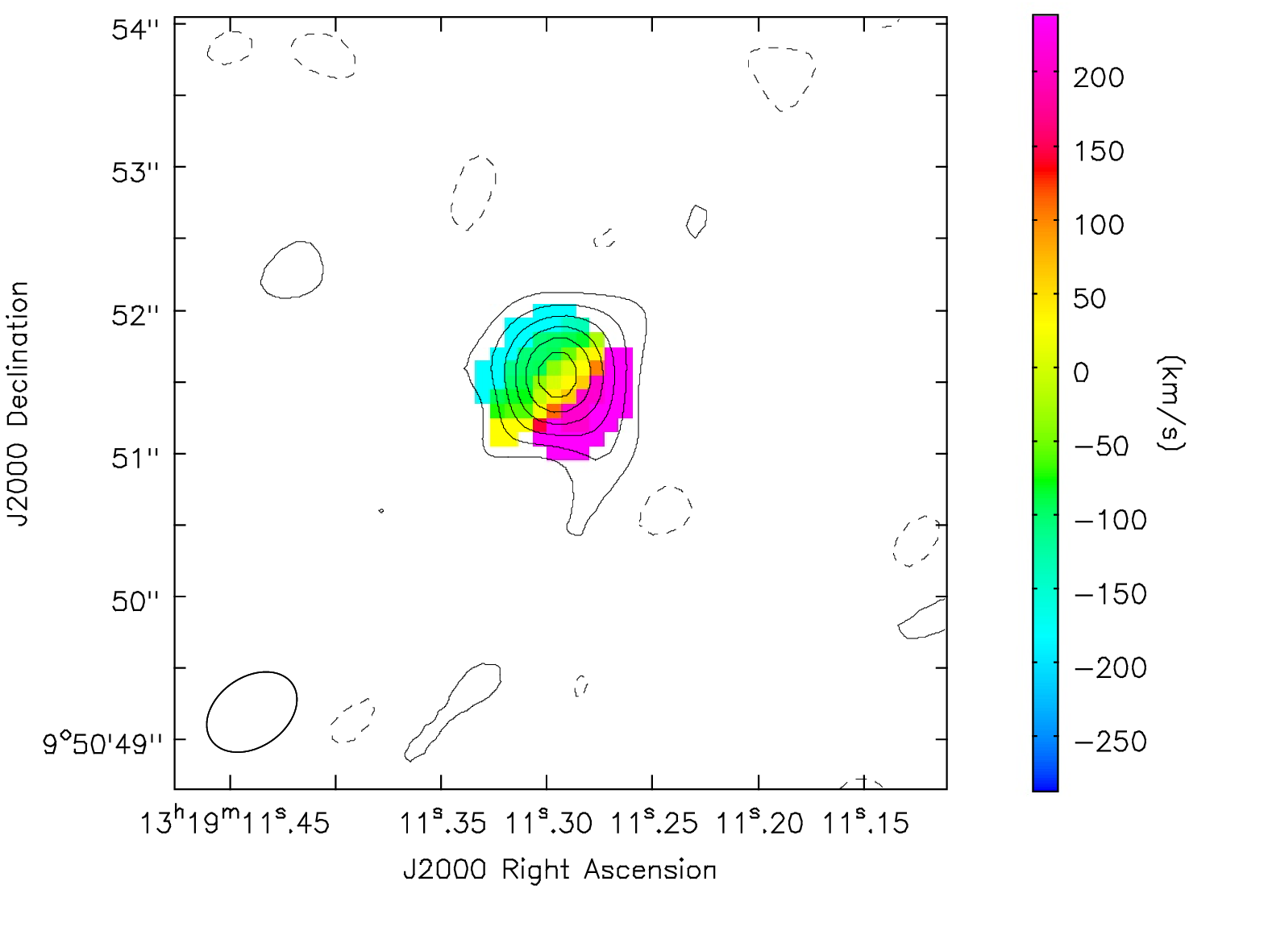}
\caption{{Left} -- The $\rm L_{FIR}$--$\rm L_{bol}$ correlation (\cite[Wang et al. 2008, 
2011]{wang08,wang11}). The 250 GHz detected z$\sim$6 quasars are plotted as black 
open squares and the arrows indicate $\rm 3\sigma$ upper limits. 
The filled four-angle and five-angle stars are 
the average values for the optically faint ($\rm m_{1450}\geq20.2$) and 
bright ($\rm m_{1450}<20.2$) 250 GHz detections, respectively, and the open ones 
shows the stacking average of the non-detections.
The local IR and PG quasars from \cite[Hao
et al. (2005)]{hao05} are plotted as cross and diamonds with
arrows denoting upper limits in $\rm L_{FIR}$. The dashed line 
represent the linear regression of $\rm log(L_{FIR})=0.45log(L_{bol})+6.62$ derived with the local 
IR quasars and the (sub)mm detected quasars at 
high redshift (\cite[Wang et al. 2011]{wang11}), and the dotted line shows the 
relationship of $\rm log(L_{FIR})=0.66log(L_{bol})+2.8$ from the local PG 
quasars(\cite[Wang et al. 2008]{wang08}).  
{Right} -- ALMA [C II] line velocity-integrated map (contours) and 
intensity-weighted velocity map (color) of
the z=6.132 quasar J1319+0950 (Wang et al. 2012, in prep.). The contours are [-2, 2, 3,
4, 6, 8, 10, 12] $\rm \times 0.18 Jy\,beam^{-1}\,km\,s^{-1}$.
The [C II] line emission is marginally resolved by the $\rm 0.7''\times0.5''$ beam, with
a source size of $\rm (0.76\pm0.05)''\times(0.74\pm0.04)''$ (deconvolved size 
of $\rm (0.6\pm0.1)''\times(0.3\pm0.2)''$, fitted using the IMFIT package in CASA).
The velocity map shows clear velocity gradient along the southwest-northeast direction. 
}
\label{fig1}
\end{figure}


\begin{thebibliography}{}
\bibitem[Bertoldi et al.(2003a)]{bertoldi03a} 
{Bertoldi, F., Carilli, C. L., Cox, P. et al.} 2003a, \textit{A\&A}, 406, L55

\bibitem[Bertoldi et al.(2003b)]{bertoldi03b} 
{Bertoldi, F., Cox, P., Neri, R. et al.} 2003b, \textit{A\&A}, 409, L47

\bibitem[Carilli et al.(2001)]{carilli01} 
{Carilli, C.L., Bertoldi, F., Rupen, M.P. et al.} 2001, \textit{ApJ}, 555, 625

\bibitem[Carilli et al.(2007)]{carilli07} 
{Carilli, C.L., Neri, R., Wang, R. et al.} 
2007, \textit{ApJ}, 666, L9

\bibitem[Fan et al.(2006)]{fan06} 
{Fan, X., Strauss, M.A., Richards, G.T., et al.} 2006, \textit{AJ}, 131, 1203

\bibitem[Hao et al. (2005)]{hao05} 
{Hao, C.N., Xia, X.Y., Mao, S., Wu, H., Deng, Z.G.} 2005, \textit{ApJ}, 625, 78

\bibitem[Jiang et al. (2009)]{jiang09}
{Jiang, L., Fan, X., Bian, F. et al.} 2009, \textit{AJ}, 138, 305

\bibitem[Kennicutt (1998)]{kennicutt98} 
{Kennicutt, R. C.} 1998, \textit{ARA\&A}, 36, 189

\bibitem[Kurk et al. (2007)]{kurk07} 
{Kurk, J.D., Walter, F., Fan, X. et al.} 2007, \textit{ApJ}, 669, 32

\bibitem[Maiolino et al.(2005)]{maiolino05} 
{Maiolino, R., Cox, P., Caselli, P. et al.} 2005, \textit{A\&A}, 440, L51

\bibitem[Maiolino et al.(2009)]{maiolino09} 
{Maiolino, R., Caselli, P., Nagao, T., Walmsley, M., De Breuck, C., Meneghetti, M.} 
2009, \textit{A\&A}, 500, L1

\bibitem[Maiolino et al.(2012)]{maiolino12}
{Maiolino, R., Gallerani, S., Neri, R. et al.} 2012, \textit{MNRAS}, 425, L66

\bibitem[Mortlock et al. 2009]{mortlock09}
{Mortlock, D.J., Patel, M., Warren, S.J. et al.} 2009, \textit{A\&A}, 505, 97

\bibitem[Omont et al. (2001)]{omont01} 
{Omont, A., Cox, P., Bertoldi, F. et al.} 2001, \textit{A\&A}, 374, 371

\bibitem[Omont et al. (2003)]{omont03} 
{Omont, A., Beelen, A., Bertoldi, F., McMahon, R. G., 
Carilli, C. L., \& Isaak, K. G.} 2003, \textit{A\&A}, 398, 857

\bibitem[Petric et al.(2003)]{petric03} 
{Petric, A. O., Carilli, C. L., Bertoldi, F. et al.} 2003, \textit{AJ}, 126, 15

\bibitem[Priddey et al. (2003)]{priddey03} 
{Priddey, R. S., Isaak, K. G., McMahon, R. G., \& Omont, A.} 2003, \textit{MNRAS}, 339, 1183

\bibitem[Riechers et al.(2006)]{riechers06} 
{Riechers, D.A., Walter, F., Carilli, C.L. et al.} 2006, \textit{ApJ}, 650, 604

\bibitem[Riechers et al.(2009)]{riechers09} 
{Riechers, D.A., Walter, F., Bertoldi, F. et al. } 2009, \textit{ApJ}, 703, 1338

\bibitem[Stacey et al. (2010)]{stacey10}
{Stacey, G.J., Hailey-Dunsheath, S., Ferkinhoff, C. et al.} 2010, \textit{ApJ}, 724, 957

\bibitem[Venemans et al. (2012)]{Venemans12}
{Venemans, B.P., McMahon, R.G., Walter, F. et al.}
 2012, \textit{ApJ}, 751, L25

\bibitem[Wagg et al. (2012)]{wagg12}
{Wagg, J., Wiklind, T., Carilli, C.L. et al.} 2012, \textit{ApJ}, 752, L30

\bibitem[Walter et al.(2003)]{walter03} 
{Walter, F., Bertoldi, F., Carilli, C. et al} 2003, Nature, 424, 406

\bibitem[Walter et al.(2004)]{walter04} 
{Walter, F., Carilli, C. L., Bertoldi, F. et al.} 2004, \textit{ApJ}, 615, L17

\bibitem[Walter et al.(2009)]{walter09} 
{Walter, F., Riechers, D., Cox, P. et al.} 2009, \textit{Nature}, 457, 699

\bibitem[Wang et al.(2008)]{wang08} 
{Wang, R., Carilli, C.L., Wagg, J. et al.} 2008, \textit{ApJ}, 687, 848

\bibitem[Wang et al.(2010)]{wang10} 
{Wang, R., Carilli, C.L., Neri, R. et al.} 2010, \textit{ApJ}, 714, 699

\bibitem[Wang et al.(2011)]{wang11}
{Wang, R., Wagg, J., Carilli, C.L. et al.}
2011a, \textit{AJ}, 142, 101

\bibitem[Willott et al.(2007)]{willott07}
{Willott, C.J., Delorme, P., Omont, A. et al.} 2007, \textit{AJ}, 134, 2435

\end{thebibliography}
\end{document}